\documentclass[10pt,preprint,aps,superscriptaddress]{revtex4-1}
\usepackage[latin9]{inputenc}
\usepackage{color}
\usepackage{amsmath}
\usepackage{amssymb}
\usepackage{graphicx}
\usepackage{subscript}
\usepackage[unicode=true,
 bookmarks=true,bookmarksnumbered=false,bookmarksopen=false,
 breaklinks=false,pdfborder={0 0 1},backref=false,colorlinks=false]
 {hyperref}
\hypersetup{pdftitle={quantum state preparation}}



\begin{document}
\title{\textcolor{black}{Proton-electron mass ratio by high-resolution optical spectroscopy of ion ensembles in the resolved-carrier regime}}
\author{I. Kortunov}
\affiliation{Institut für Experimentalphysik, Heinrich-Heine-Universität Düsseldorf,
40225 Düsseldorf, Germany}
\author{S. Alighanbari}
\affiliation{Institut für Experimentalphysik, Heinrich-Heine-Universität Düsseldorf,
40225 Düsseldorf, Germany}
\author{M.~G. Hansen}
\affiliation{Institut für Experimentalphysik, Heinrich-Heine-Universität Düsseldorf,
40225 Düsseldorf, Germany}
\author{G.~S.~Giri}
\affiliation{Institut für Experimentalphysik, Heinrich-Heine-Universität Düsseldorf,
40225 Düsseldorf, Germany}
\author{V.~I. Korobov}
\affiliation{Bogoliubov Laboratory of Theoretical Physics, Joint Institute for
Nuclear Research, 141980 Dubna, Russia}
\author{S. Schiller}
\email[Corresponding Author, e-mail:~]{step.schiller@hhu.de}
\affiliation{Institut für Experimentalphysik, Heinrich-Heine-Universität Düsseldorf,
40225 Düsseldorf, Germany}
\begin{abstract}

\textbf{Optical spectroscopy in the gas phase is a key tool to elucidate
the structure of atoms and molecules and of their interaction with external fields. The line resolution is usually
limited by a combination of first-order Doppler broadening due to
particle thermal motion and of a short transit time through the excitation
beam. For trapped particles, suitable laser cooling techniques can
lead to strong confinement (Lamb-Dicke regime, LDR) and thus to optical
spectroscopy free of these effects. For non-laser coolable spectroscopy
ions, this has so far only been achieved when trapping one or two
atomic ions, together with a single laser-coolable atomic
ion \cite{Rosenband2007,Chou2011}.  Here we show that one-photon
optical spectroscopy free of Doppler and transit broadening can also
be obtained with more easily prepared ensembles of ions, if performed
with mid-infrared radiation. We demonstrate the method on molecular
ions. We trap approximately 100 molecular
hydrogen ions (HD\textsuperscript{+}) within a
Coulomb cluster of a few thousand laser-cooled atomic ions and perform
laser spectroscopy of the fundamental vibrational transition. Transition frequencies were
determined with lowest uncertainty of $3.3\times10^{-12}$
fractionally. As an application, we determine the proton-electron
mass ratio by matching a precise \emph{ab initio} calculation with the measured vibrational frequency.} 
\end{abstract}
\maketitle

The pursuit of increasingly higher resolution in spectroscopy is of
fundamental importance in the field of atomic and molecular physics.
Techniques such as nonlinear spectroscopy, particle trapping, laser
cooling, buffer gas cooling, and improved microwave and laser sources
have permitted continuous progress in resolution over more than half
a century, resulting in major advances in understanding of radiation-matter
interactions and in controlling quantum systems, especially in the
optical domain. One key approach to ultra-high (Doppler-free) resolution is the Lamb-Dicke
regime (LDR), achieved by confinement of particles, at least in one
spatial dimension, to a range significantly smaller than the wavelength
of the spectroscopy radiation \cite{Dicke1953a}. Historically, spectroscopy
in this regime was first introduced for radiofrequency and microwave
spectroscopy, for diverse neutral and charged particles. For hyperfine
structure spectroscopy of trapped atomic ion clouds, it was shown
already in the earliest studies \cite{Major1973} that spectral lines
with extremely high quality factors are achievable if the cloud radius
is smaller than the wavelength.

For visible radiation ($\lambda\simeq0.5\,\mu{\rm m}$), the LDR imposes
greater experimental challenges, but is today achieved in many experiments.
LDR optical spectroscopy has already led to spectacular advances in
atomic clock performance (for both atomic ions and neutral atoms)
\cite{Rosenband2008,Derevianko2011,Poli2013,Ludlow2015-1} and is
also an enabling technique for atom-based quantum information processing,
cold-atom many-body system studies, and quantum simulations.  

For laser-coolable atomic ions, optical LDR spectroscopy has been demonstrated
for ion numbers ranging from 1 to 18, arranged in a string \cite{Diedrich1989,Lechner2016}.
For some atomic ions species and most molecular ion species, direct
laser cooling is impractical or not possible. One then resorts to
the technique of sympathetic cooling. LDR spectroscopy of a single
and of two sympathetically cooled atomic ions has been demonstrated,
using quantum logic spectroscopy \cite{Schmidt2005,Rosenband2007,Chou2011}.

For molecules, the LDR has very recently been achieved in a few different
configurations: (i) one-photon rotational spectroscopy of molecular
ion ensembles at 1.3~THz transition frequency \cite{Alighanbari2018,Alighanbari2020}, (ii) two-photon Raman rotational spectroscopy of a single molecular ion \cite{Chou2020}, (iii) two-photon Raman vibrational spectroscopy (25~THz) of laser-photoassociated diatomic neutral molecule ensembles in an optical lattice trap \cite{Kondov2019}, (iv) two-photon vibrational
spectroscopy of molecular ion ensembles, using counter-propagating
waves of close wavelength $(\lambda\simeq1.4\,\mu{\rm m})$ \cite{Patra2020}.
Some of the line resolutions  (defined as transition frequency divided
by full width at half maximum) of up to $8\times10^{11}$ achieved
in these works \cite{Kondov2019} have surpassed the highest values
achieved with room temperature gaseous ensembles and on molecular
beams (see Methods). 

In this letter, we demonstrate that it is possible to perform ultra-high-resolution
\emph{one-photon} mid-infrared optical spectroscopy of cold ions,
on comparatively large and easily prepared \emph{ensembles} of ions.
While this is here demonstrated on sympathetically cooled ions, it
is expected to work also with ions that are laser cooled directly.
The present demonstration is a powerful extension of the previously
introduced technique TICTES (Trapped Ion Cluster Transverse Excitation
Spectroscopy) from rotational ($\simeq1$~THz) to vibrational spectroscopy
($\simeq50$~THz) \cite{Alighanbari2018,Alighanbari2020}. The technique
operates on ion ensembles in a macroscopic linear ion trap and with
standard Doppler laser cooling. The crucial aspect is the use of a string-like
spatial arrangement of the spectroscopy ions and a spectroscopy radiation
having a relatively large wavelength (here 5.1~$\mu$m) and propagated
at right angle to the ion trap axis. The technique is to be contrasted
with quantum-logic spectroscopy \cite{Schmidt2005,Wolf2016a,Chou2020}. Although the latter is powerful and likely extendable to vibrational spectroscopy
of molecular ions, its ion preparation technique and the actual spectroscopy
procedures are complex and so far restricted to a few specialized
research groups world-wide.

 In TICTES, two species of ions are simultaneously trapped in a
linear ion trap, where one species (LC) is laser-cooled, the other,
``target'' species (SC) is sympathetically cooled. The corresponding
ion numbers are assumed to be $N_{{\rm LC}}\gg N_{{\rm SC}}\gg1$.
 If the two species have charges ($q$) and masses ($m$) such that
$q_{{\rm SC}}^{2}/m_{{\rm SC}}>q_{{\rm LC}}^{2}/m_{{\rm LC}}$, the
species SC will be confined to the region near the trap axis.  
Here, we focus on a configuration in which the SC ions form an ion
string embedded in the LC ion cluster. The dynamics of such a two-species
cluster at finite ion temperature has been previously analyzed by
means of molecular dynamics simulations \cite{Alighanbari2018}. The
time-averaged spatial distribution of SC ions has a finite width in
radial direction, $\Delta\rho$. For ensembles of HD\textsuperscript{+}
ions cooled by beryllium ions to a temperature of $T\simeq10\,{\rm mK}$
and arranged in a string-like configuration, $\Delta\rho\simeq2\,\mu{\rm m}$
was found in the simulations, for our trap parameters. A Doppler-free
carrier signal with frequency unaffected by ion dynamics was predicted
to occur for a spectroscopy wave ($\lambda$) irradiation direction transverse
to the trap axis (radial direction) \cite{Zhang2008}, assuming that
the intrinsic linewidth of the transition is sufficiently narrow.
Both numerically and within an approximate analytical model it was
furthermore predicted that the strength of the carrier spectroscopy
signal decreases continuously from near-unity if $\lambda\gg\lambda_{c}=2\pi\Delta\rho$
(LDR) to zero for $\lambda\ll\lambda_{c}$. At $\lambda=\lambda_{c}$
the carrier signal is 0.5. While in the first experimental investigation
the regime $\lambda\gg\lambda_{c}$ was studied \cite{Alighanbari2018},
in this work we explore whether a carrier signal is still observable
for a small spectroscopy wavelength, $\lambda\simeq5.1\,\mu{\rm m}<\lambda_{c}$.
At this wavelength, the carrier signal strength is predicted to be
less than 0.02 \cite{Alighanbari2018}. We note that in microwave
spectroscopy of trapped atomic ion ensembles, ultra-high resolution
of the carrier transition has been observed in the non-LD regime.
\cite{Major1973,Lakkaraju1982,Cutler1985,Prestage1990,Fisk1997}.

Wavelengths of 5~$\mu$m and longer permit addressing  fundamental
and/or first overtone vibrational transitions of most molecular ions,
except for hydrides. Such vibrational transitions have natural linewidths
in the range from 10~Hz in heteronuclear molecules to nano-Hz in
homonuclear diatomics. Attainment of the resolved carrier regime can
in principle take advantage of these small linewidths to provide e.g.
a high spectral resolution of hyperfine structure, a high sensitivity
in the study of frequency shifts caused by interaction with externally
applied fields, and precise measurements of transition frequencies.

\begin{figure}[t]
\centering{}
\includegraphics[width=1\columnwidth]{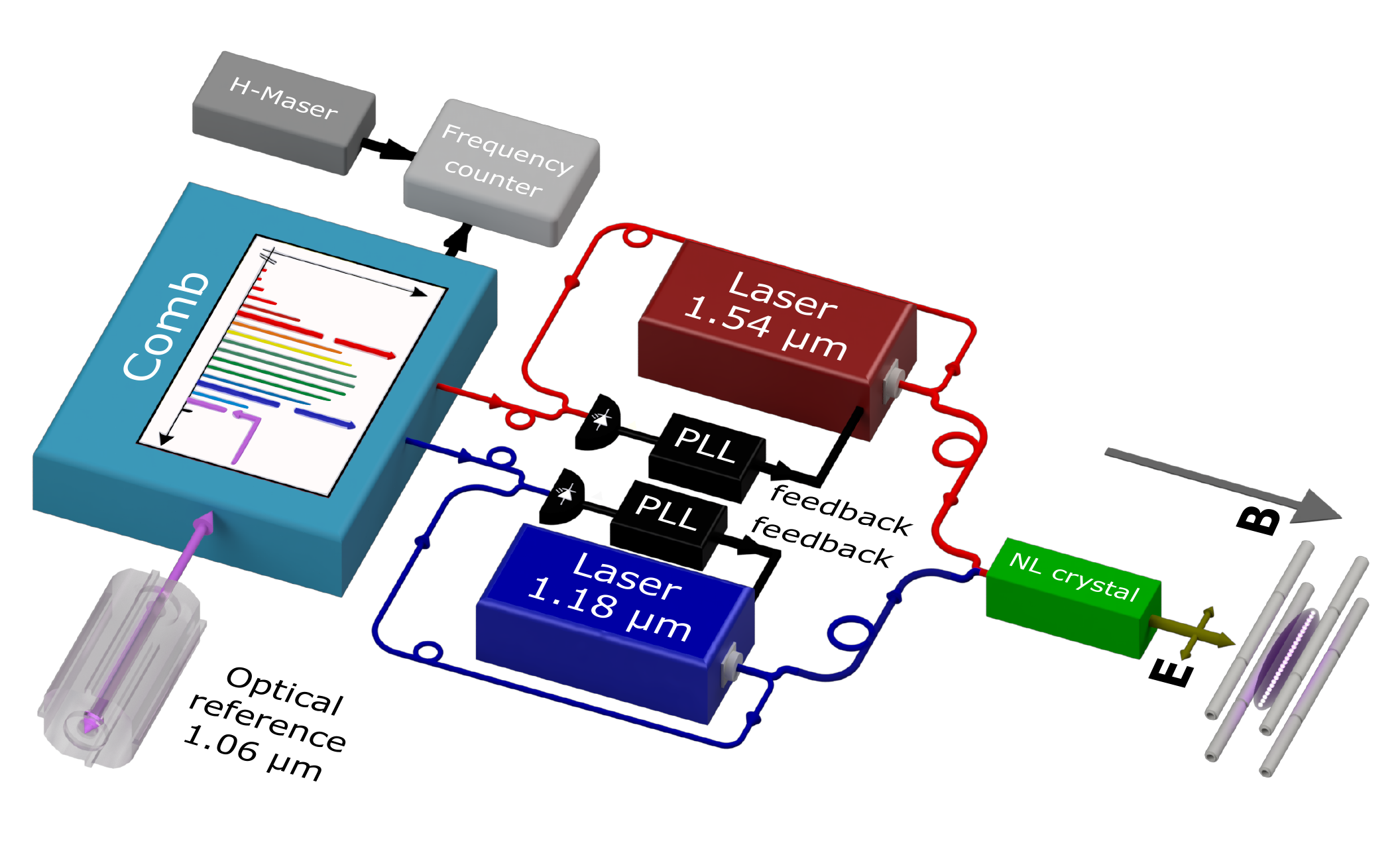}\caption{\label{fig:Schematic of apparatus} \textbf{Scheme of the key elements
of the apparatus.} The linear ion trap is at the right. The shape
of the two-species Be\protect\textsuperscript{+}/HD\protect\textsuperscript{+}
ion cluster is shown. The mid-infrared radiation generated by a nonlinear-optical (NL) crystal (olive arrow) propagates perpendicularly to the ion cluster's
long axis. The two lasers L1, L2 are phase-locked to two modes of
the femtosecond frequency comb (red and blue arrows in the inset diagram
on the comb). The comb is phase-locked to a continuous-wave laser
(1.06~$\mu$m), stabilized to an optical resonator. Only that resonator
is shown (bottom left). PLL: phase-locked loop.  }
\end{figure}

\begin{figure}[t]
\includegraphics[width=\columnwidth]{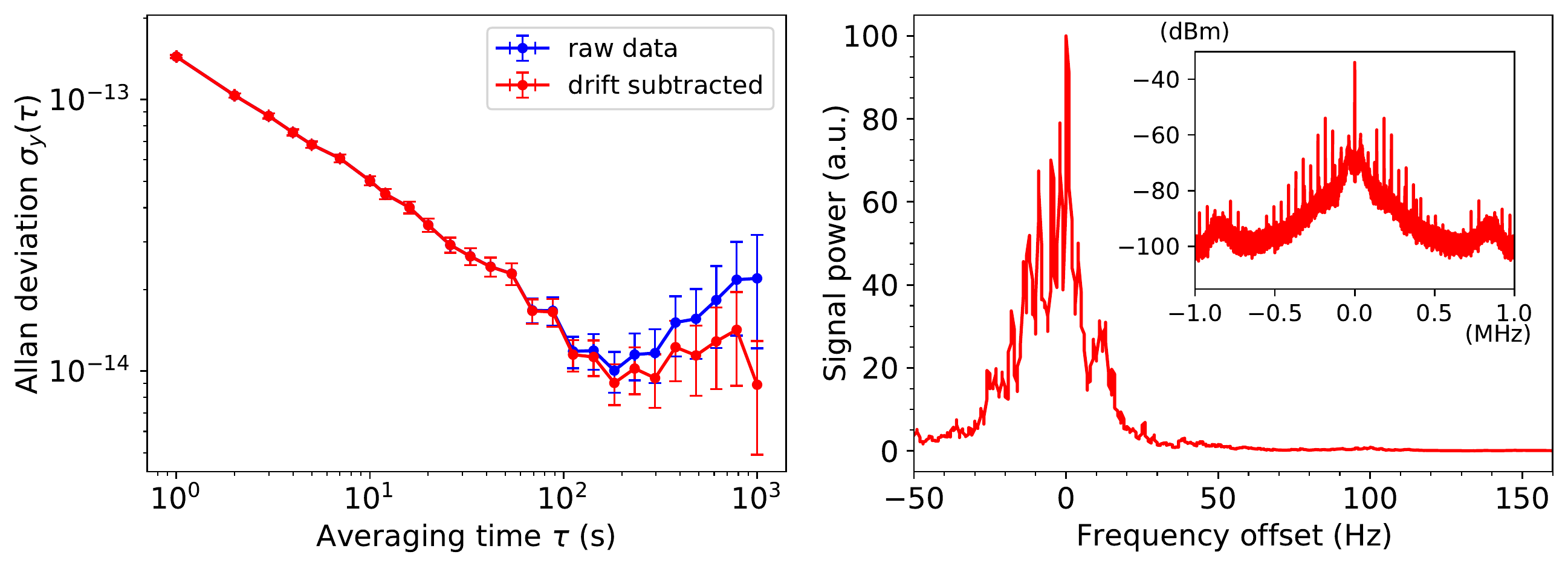}
\caption{\label{fig: Spectral properties of laser} \textbf{Spectral properties
of the mid-infrared laser source.} Left: Fractional Allan deviation of the computed frequency value of the 5.1~$\mu$m radiation. Error bar for Allan deviation is estimated as 68\%-confidence interval. Right: the heterodyne beat between a mode of the optically-stabilized frequency comb and an independent frequency-stable 1.5~$\mu$m laser (not shown in Fig.~\ref{fig:Schematic of apparatus}).
Averaging time: 1.5~min. Inset: beat on a larger frequency scale.
The observed linewidth of less than 50~Hz, together with other data
(see Methods), indicates that the mid-infrared radiation has a linewidth
of similiar value.}
\end{figure}

To achieve a thorough exploration of the technique's potential in terms of
resolution and accuracy required we developed an appropriate
laser source with ultra-narrow spectral linewidth, ultra-high long-term
absolute frequency stability, and precise frequency calibration. In
the mid-infrared spectral range such sources are not routinely available.
 Our source is based on generation of difference-frequency radiation
($\lambda\simeq5.1\,\mu{\rm m}$) from two individually frequency-stabilized,
commercial semiconductor lasers ($\lambda_{1}=1.18\,\mu{\rm m}$,
$\lambda_{2}=1.54\,\mu{\rm m}$), see Fig.~\ref{fig:Schematic of apparatus}. We compute the frequency of the mid-infrared radiation from the two
laser frequencies, that we measure using a near-infrared frequency
comb. Figure~\ref{fig: Spectral properties of laser} presents
data on the spectral purity and frequency stability of our source.
We estimate the linewidth of the $5.1\,\mu$m radiation to be less
than 100~Hz (see Methods).

Our test ion is the one-electron diatomic molecular ion ${\rm HD}^{+}$
of mass 3~u. The choice of this ion is owed to the feasibility of
\textit{ab initio} calculation of transition frequencies, so that
the  experimentally assessed spectroscopic \textit{accuracy} of the
method can also be independently verified. The \emph{ab initio} calculation
method itself has been stringently tested by a recent related experiment
in the same ion trap \cite{Alighanbari2020}. We interrogate the
electric-dipole-allowed, one-photon fundamental vibrational transition
$(v=0,\,N=0)\rightarrow$ $(v'=1,\,N'=1)$ in the ground electronic
state $(^{2}\Sigma_{g}^{+})$. $v,\,N$ denote the rotational and
vibrational quantum number, respectively.  The transition frequency
is $f\simeq58.6\,{\rm THz}$. See Methods for details.

${\rm HD}^{+}$ possesses hyperfine (spin) structure and we focus on two transition components, denoted by line~12 and line~16 (see Methods
and Supplementary Information, SI), which share the same lower hyperfine
state. The respective upper hyperfine states have the same particle
spin coupling but differ in the coupling between total particle spin
and rotational angular momentum. This results in different total angular
momenta $F'$ in the upper states. For maximum resolution we address individual Zeeman components of the
spin structure of the vibrational transition in the presence of a
small magnetic field. We specifically measure the $m_{F}=0\rightarrow m_{F}'=0$
Zeeman components, because they exhibit only a comparatively small
quadratic Zeeman shift (see Methods). Here, $m_{F}$ ($m_{F}'$) is
the projection of the total angular momenta $F$ ($F'$) on the static
magnetic field direction.

\begin{figure}[t]
\includegraphics[width=\textwidth]{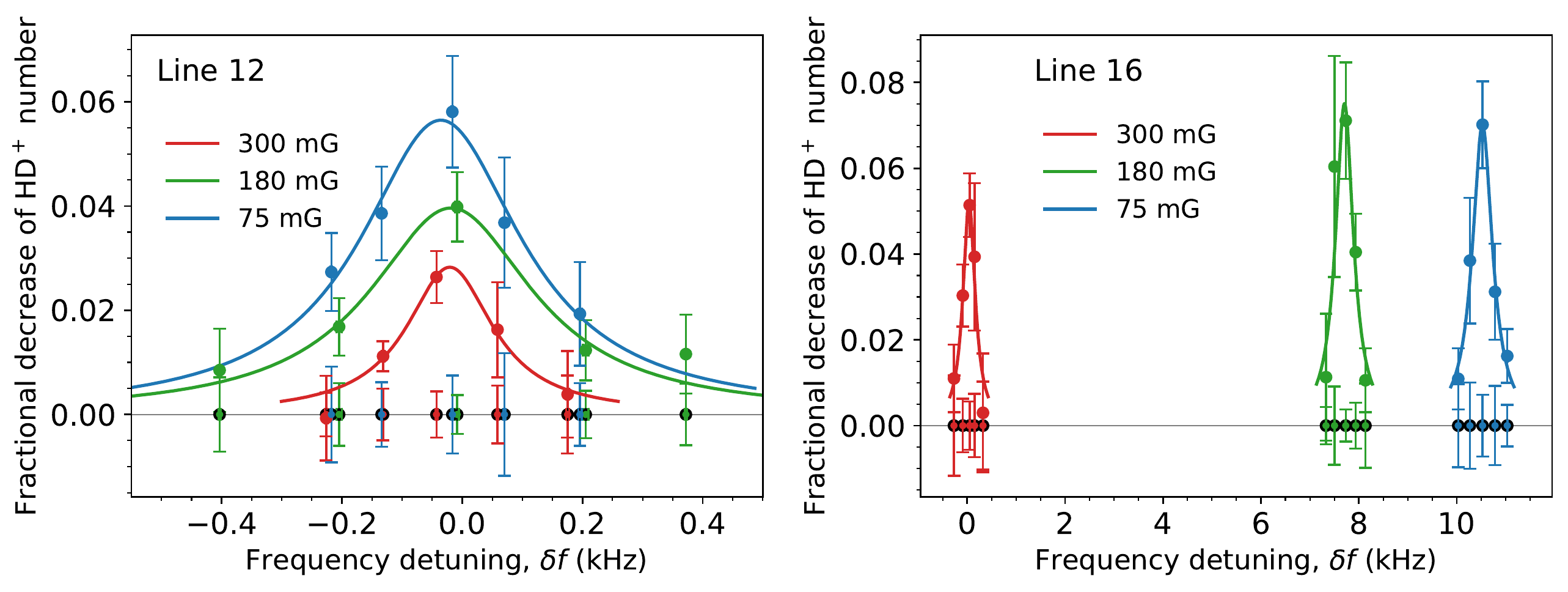}

\caption{\label{fig: Narrow Transition}\textbf{Two hyperfine components of
the fundamental rovibrational transition of HD}\protect\textsuperscript{\textbf{+}}\textbf{,
at a frequency of 58.6~THz.} Two Zeeman components $(v=0,\,N=0,\,F=2,\,m_{F}=0)\rightarrow(v'=1,\,N'=0,\,F'=1,\,m_{F}'=0)$
(line~12) and $(v=0,\,N=0,\,F=2,\,m_{F}=0)\rightarrow(v'=1,\,N'=1,\,F'=3,\,m_{F}'=0)$
(line~16) are shown, each for three values of the applied magnetic
field. The quadratic Zeeman shift is evident on line 16, while it
is not resolved on line 12. For line 16 at the lowest magnetic field
setting, the line contains also the two transitions between stretched
states ($m_{F}=\pm2\rightarrow m_{F}'=\pm3$). For each indicated
detuning $\delta f$ of the laser frequency, two sets of measurements
of HD\protect\textsuperscript{+} number decay were performed: with
(colored) and without application of spectroscopy radiation (background
decay, colored and circled black). From the mean of each set the background
decay mean was subtracted. The zero detuning frequency is arbitrary.
The linewidths are due to a combination of power broadening and spectroscopy wave linewidth. The theoretical value for the natural linewidth of the transition is approximately 3~Hz. The lines are Lorentzian fits. Each error bar represents the standard deviation of the mean. The red data points and the fit for line~12 have been shifted by 0.17~kHz for clarity.}
\end{figure}

Figure~\ref{fig: Narrow Transition} shows one Zeeman component
of line~12 and of line~16. The highest line resolutions obtained
were $3\times10^{11}$ (full linewidths smaller than 0.2~kHz). We
measured several systematic shifts. 1) Zeeman shift; we measured the
transition frequency of each line for three values of the magnetic
field $B<1\,{\rm G}$. The shifts are consistent with the \emph{ab
initio} prediction. Assuming the predicted quadratic-in-$B$ scaling,
an extrapolation to $B=0$ was performed. 2) The laser light for beryllium
ion laser cooling ($0.2-0.5$~mW at 313~nm) potentially causes a
light shift via the polarisability of the molecular ion. A measurement
showed that there is no effect at $0.2$~kHz level, and the theoretical
estimate justifies setting this shift to zero in our analysis. 3)
A shift caused by the trap RF electric field.
Measurement of the transition frequency of each line for three RF
amplitude values allowed for an extrapolation to zero amplitude. The
extrapolated values are approximately 0.3~kHz smaller than the values
at our nominal operational RF amplitude. 4) We irradiate the spectroscopy wave and the two photodissociation lasers alternatingly to avoid light shifts caused by the latter. 5) Our theoretical estimation reveals that other systematic shifts are negligible compared to the uncertainties resulting from the above determinations. See Methods for further details.

We obtain the extrapolated zero-field frequencies 
\begin{align}
f_{{\rm 12}}^{{\rm (exp)}} & =58\,605\,013\,478.03(19)_{{\rm exp}}~{\rm kHz}\,,\label{eq: f_vib,exp}\\
f_{{\rm 16}}^{{\rm (exp)}} & =58\,605\,054\,772.08(26)_{{\rm exp}}\text{ }{\rm kHz}\,.\nonumber 
\end{align}
 The indicated uncertainties result from the realized linewidths
and the achieved precision of the determination of the systematic
shifts. Thus, the lowest experimental uncertainty is $3.3\times10^{-12}$
fractionally.

The difference of the frequencies in eq.~(\ref{eq: f_vib,exp}) is
a spin-rotation splitting. Its experimental value can be compared
with the predicted splitting, eq.~(\ref{eq:theory fspin16 - fspin12}) (see Methods).
Theory and experiment agree very well within the combined uncertainty
of  0.54~kHz.

We furthermore compare the above values with \textit{ab initio} values
$f_{i}^{{\rm (theor)}}$, computed using the approach described in
Methods. Using current (2018) Committee on Data for Science and Technology
(CODATA~2018) values of the fundamental constants \cite{Tiesinga2019}
and their uncertainties (case I) results in 
\begin{align}
f_{{\rm 12}}^{{\rm (theor)}} & =58\,605\,013\,477.8(5)_{{\rm theor,QED}}(8)_{{\rm theor,spin}}(13)_{{\rm CODATA2018}}\,{\rm kHz\,,}\nonumber \\
f_{{\rm 16}}^{{\rm (theor)}} & =58\,605\,054\,771.6(5)_{{\rm theor,QED}}(9)_{{\rm theor,spin}}(13)_{{\rm CODATA2018}}\,{\rm kHz\,.}\label{eq:f_theor_total}
\end{align}
For both frequencies, the indicated uncertainties  are $(0.8,1.5,2.2)\times10^{-11}$
in fractional terms. The first uncertainty of both frequencies, 0.5~kHz,
is an estimate of the unevaluated QED contributions. It has been reduced
by a factor~42 in theoretical work spanning the last 9~years. If
for $m_{e}$, $m_{p}$, $m_{d}$ the most accurate Penning trap mass
values \cite{Koehler2015,Heisse2019,Rau2020} are used (case II),
instead of their CODATA 2018 values, the predictions are shifted by
+1.5~kHz, and the last uncertainty contribution reduces to 1.1~kHz.
Our experimental values are consistent with both predictions
I,~II within the combined uncertainties.

Any normalized linear combination of the two experimental vibrational
frequencies, with respective theoretical spin structure contributions
subtracted, yields the spin-averaged vibrational frequency,
\begin{equation}
f_{{\rm spin-avg}}^{({\rm exp})}=b_{12}(f_{{\rm 12}}^{({\rm exp})}-f_{{\rm spin},12}^{({\rm theor})})+(1-b_{12})(f_{{\rm 16}}^{({\rm exp})}-f_{{\rm spin},16}^{({\rm theor})})\,.\label{eq: f_spin-avg exp formula}
\end{equation}
We may choose the weight $b_{12}$ in this composite frequency so
that the total spin theory uncertainty is minimized \cite{Alighanbari2020}.
However, we find that the uncertainty is minimized over a wide range
of $b_{12}$ values between 0 and 1, without any significant reduction
compared to that of $f_{{\rm spin,12}}^{{\rm (theor)}}$ and $f_{{\rm spin,16}}^{{\rm (theor)}}$.
For $b_{12}\simeq0.5$, 
\begin{align}
f_{{\rm spin-avg}}^{{\rm (exp)}} & =58\,605\,052\,164.24(16)_{{\rm exp}}(85)_{{\rm theor,spin}}\,{\rm kHz\,.}\label{eq:fspin-avg-exp}
\end{align}
The value $f_{{\rm spin-avg}}^{{\rm (exp)}}$ agrees with the prediction
$f_{{\rm spin-avg}}^{{\rm (theor)}}$, eq.~(\ref{eq:f_spin-avg}),
both for case~I and II. The combined uncertainty of experimental
and predicted values is $2.9\times10^{-11}$. This realizes a test of three-body quantum physics with state-of-the-art precision, and limited by the uncertainties of of the mass values.

\begin{figure}[t]
\includegraphics[width=\columnwidth]{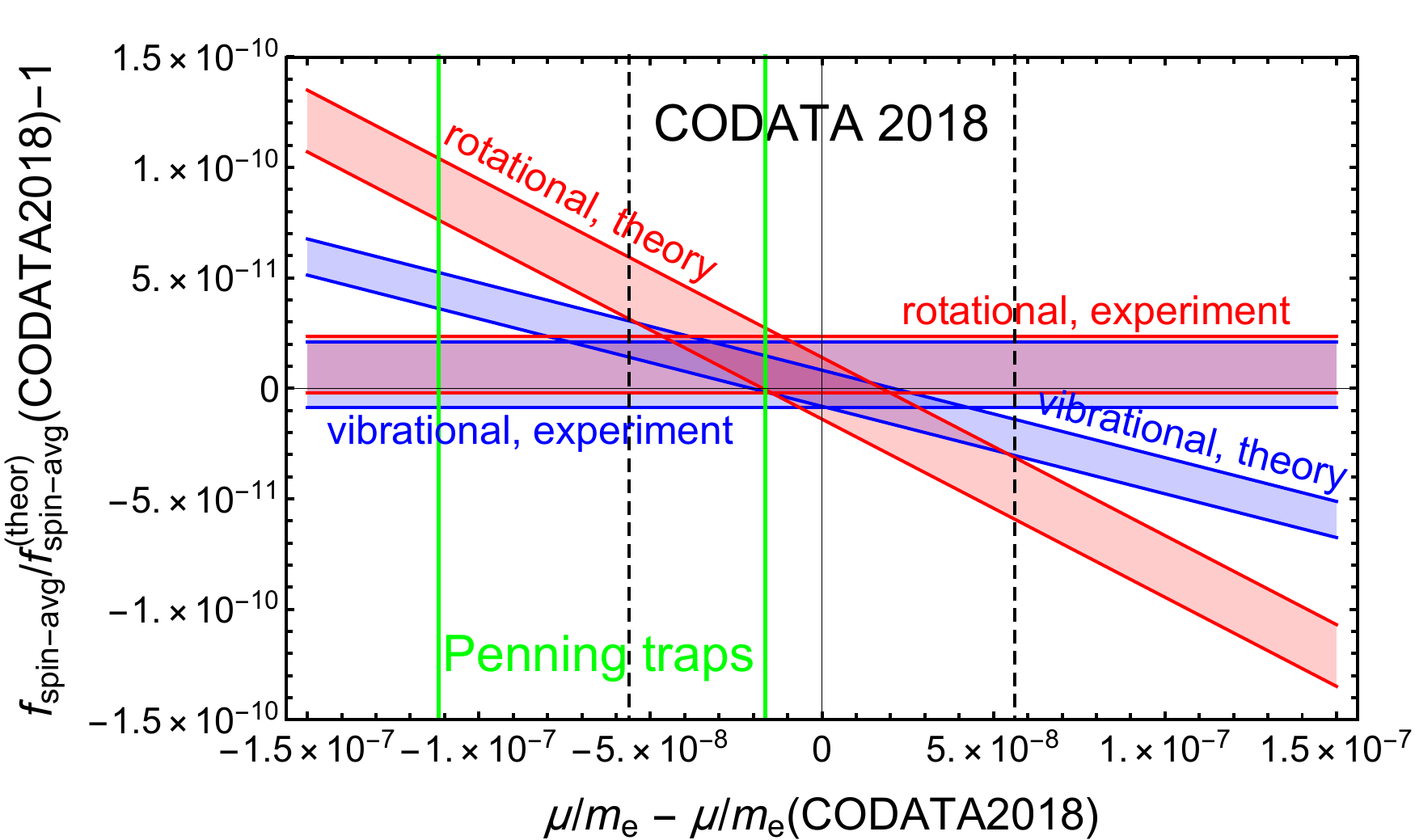}\caption{\label{fig:Plot_of_reducedmass}\textbf{Determinations of the ratio
of reduced nuclear mass to electron mass $\mu/m_{{\rm e}}=m_{{\rm e}}^{-1}m_{{\rm p}}m_{{\rm d}}/(m_{{\rm p}}+m_{{\rm d}})\simeq1223.9$.}
 Theoretical predictions (tilted bands) and experimental results
(horizontal bands) are compared. Blue: this work. Red: rotational
transition \cite{Alighanbari2020}. The widths of the bands represent
twice the uncertainties. For the prediction bands, the uncertainty
includes the uncertainties of \emph{ab initio} theory and the CODATA
2018 uncertainties of the fundamental constants, excluding $\mu/m_{{\rm e}}$.
For the experimental bands, the uncertainties include the uncertainties
of experiment and of the spin contribution correction (see eq.~(\ref{eq: f_spin-avg exp formula})).
For the rotational transition measurement, the latter is negligible.
Black dashed: CODATA 2018 $\pm1\,\sigma$ uncertainty range. Green:
$\pm1\,\sigma$ uncertainty range for the value $\mu/m_{{\rm e}}$
computed from the results of refs.~\cite{Koehler2015,Heisse2019,Rau2020}.}
\end{figure}

The fundamental vibrational frequency $f_{\rm spin-avg}^{\rm theor}$ depends on the nuclear masses
dominantly via the reduced nuclear mass $\mu=m_{{\rm p}}m_{{\rm d}}/(m_{{\rm p}}+m_{{\rm d}})$,
being closely proportional to $R_{\infty}\sqrt{\mu/m_{e}}$. We may
therefore determine the ratio $\mu/m_{{\rm e}}$ by requiring $f_{{\rm {\rm spin-avg}}}^{{\rm (theor)}}(\mu/m_{e})=f_{{\rm {\rm spin-avg}}}^{{\rm (exp)}}$,
\begin{equation}
\mu/m_{{\rm e}}=1\,223.899\,228\,668\,(7)_{{\rm exp}}(20)_{{\rm theor,QED}}(37)_{{\rm theor,spin}}(3)_{{\rm CODATA2018}}.\label{eq:=0000B5overme from experiment and theory}
\end{equation}
The last uncertainty contribution is due to that of the nuclear charge
radii and the Rydberg constant $R_\infty$, and the total fractional uncertainty
is $u_{{\rm r}}=3.5\times10^{-11}$. As shown in Fig.~\ref{fig:Plot_of_reducedmass},
the value eq.~(\ref{eq:=0000B5overme from experiment and theory})
is consistent with the values from (i) CODATA 2018, (ii) recent Penning
trap measurements of $m_{{\rm e}}$, $m_{{\rm p}}$, $m_{{\rm d}}$
and $m_{{\rm d}}/m_{{\rm p}}$ \cite{Sturm2014,Heisse2019,Fink2020,Rau2020},
and (iii) our HD\textsuperscript{+} experimental rotational frequency
and its theory \cite{Alighanbari2020}. The consistency between our
vibrational and rotational values represents a direct test of the
correctness of our QED and spin theory for this molecule, where today's
uncertainties of the fundamental constants do not enter at a relevant
level.

Since the CODATA uncertainty of $\mu_{{\rm p}}=m_{{\rm p}}/m_{{\rm e}}$
contributes most to the CODATA 2018 uncertainty of $f_{{\rm {\rm spin-avg}}}^{{\rm (theor)}}$,
we may alternatively fit $m_{{\rm p}}/m_{{\rm e}}$. In this case,
for $m_{{\rm d}}/m_{{\rm p}}$ we use the mean of the two recent precise
values measured with Penning traps \cite{Fink2020,Rau2020}. For the
remaining fundamental constants we use the CODATA 2018 values. We
obtain 
\begin{align}
m_{{\rm p}}/m_{{\rm e}} & =1\,836.152\,673\,384\,(11)_{{\rm exp}}(31)_{{\rm theor,QED}}(55)_{{\rm theor,spin}}(12)_{{\rm CODATA2018,Fink-Rau}}\,,\label{eq: me/mp from experiment and theory}
\end{align}
with total fractional uncertainty $u_{{\rm r}}=3.5\times10^{-11}$.
The value is in agreement with other recent precision measurements:

(1) $1\,836.152\,673\,374\,(78)_{{\rm exp}}$, obtained from Penning
trap determinations \cite{Koehler2015,Heisse2019},

(2) $1\,836.152\,673\,449\,(24)_{{\rm exp}}(25)_{{\rm theor,QED}}(13)_{{\rm CODATA2018,Fink}}$
obtained from combining rotational spectroscopy of HD\textsuperscript{+}
and one Penning trap measurement of $m_{{\rm d}}/m_{{\rm p}}$ \cite{Fink2020},
and

(3) $1\,836.152\,673\,349\,(71)$ obtained from a HD\textsuperscript{+}
high-overtone vibrational frequency, its theory and CODATA2018 constants
\cite{Patra2020}. 

We note that our result, eq.~(\ref{eq: me/mp from experiment and theory}),
is limited in precision by the hyperfine structure theory. In the
future this limitation may be overcome by an improved theory \cite{Korobov2020}.

In conclusion, we demonstrated a one-photon optical spectroscopy method
for ensembles of molecular ions that achieved a more than  $10^{4}$
higher fractional resolution \cite{Bressel2012} and  a 400-fold
higher fractional accuracy than previously \cite{Bressel2012,Biesheuvel_2016}. We improved the absolute accuracy in determining a molecular-ion
hyperfine splitting by optical spectroscopy by  a factor
400. We also achieved an uncertainty $3.3\times10^{-12}$ in the
experimental determination of a vibrational transition frequency of
the present test ion, by measurement and theoretical evaluation of
the systematic shifts. Because the employed test molecule is not particularly
insensitive to external perturbations, we expect that similar ($10^{-12}$)
inaccuracy levels should be achievable for numerous other molecular
ion species. An independent test of the accuracy was possible by
comparing our experimental composite transition frequency with the
\emph{ab initio} calculation. We found agreement at the $2.9\times10^{-11}$~level.
We also showed the usefulness of the technique for the field of precision
physics, by deriving values of fundamental mass ratios with uncertainties
close to those of the most precise measurements \cite{Sturm2014,Heisse2019,Alighanbari2020,Patra2020}. A possible
further exploitation of the technique is the study of molecular ions
for tests of the equivalence principle \cite{Schiller2005}.
We believe that the present technique is applicable also to electric
quadrupole vibrational transitions of homonuclear diatomic ions \cite{Germann2014}, which are relevant for the above studies \cite{Schiller2014,Karr2014}.

While we employed a light coolant atomic ion in the present demonstration, coolant atomic ions of larger mass are suitable for co-trapping
and sympathetic cooling molecular ions of correspondingly larger mass.
With a coolant ion mass upper bound of 160~u a vast variety of singly
charged molecular ions can be covered. The experimental configuration
and employed techniques are comparatively accessible and should lend
themselves to adoption by the spectroscopy community. Future work
will explore the resolution and accuracy limits of this novel technique,
and applicability to single-species ensembles.

\clearpage{}

\clearpage{}

\clearpage{}

\section{Methods}
\begin{NoHyper}
\subsection{Experimental Apparatus}

The ion trap apparatus used in the present work (Fig.~\ref{fig:Schematic of apparatus})
has been described previously \cite{Alighanbari2018,Alighanbari2020}.
The ion trap exhibits a minimum distance between diagonally opposing
electrodes of 8.6~mm. We have newly developed the laser system, which consists of two continuous-wave high-power diode lasers (L1, L2)
emitting at 1.18~$\mu$m and 1.54~$\mu$m, whose radiation is mixed
in a periodically-poled LiNbO\textsubscript{3} crystal to produce
the difference-frequency wave having a wavelength of 5.1~$\mu$m.

L1 and L2 are each phase-locked to a femtosecond fiber frequency comb,
which is itself phase-locked to a 1.06~$\mu$m laser. We stabilize this laser to an ultra-low-expansion glass cavity~{[}37{]},
representing the optical flywheel reference of the whole system. Our
phase locking scheme allows flexible tuning of the mid-infrared radiation,
limited by tunability of the pump lasers.

The three optical beats required to implement the three phase locks
are continuously monitored with spectrum analyzers. The beats exhibit
RF linewidths below 1~Hz. To independently verify the quality of
the frequency comb stability, we show in Fig.~\ref{fig: Spectral properties of laser}~(right)
the beat between a fourth diode laser (1.56~$\mu$m) stabilized to
its own reference cavity~{[}36{]} (both not shown in Fig.~\ref{fig:Schematic of apparatus})
and an appropriate comb mode. This wavelength 1.56~$\mu$m is significantly
different from wavelength of the phase-locked comb mode, 1.06~$\mu$m
and therefore the beat is a sensitive monitor of comb frequency noise.
The full-width-at-half-maximum linewidth of the beat is less than
50~Hz. We infer that the linewidths of the lasers L1, L2 are also
less than 50 Hz. The linewidth of the mid-infrared radiation is probably
significantly less than 100~Hz because there is common-mode frequency
noise in L1 and L2, originating from the 1.06-$\mu$m-laser's frequency
instability.

We determine the absolute optical frequencies $f_{1},\,f_{2}$ of L1, L2 in real time by measuring the repetition rate of the frequency comb
and the carrier envelope offset frequency $f_{{\rm ceo}}$. A hydrogen maser provides the reference frequency for the frequency counter. The frequency of the mid-infrared radiation is computed as $f_{0}=f_{1}-f_{2}$.
Note that $f_{{\rm ceo}}$ drops out. The absolute frequency measurement
uncovers the slow frequency drift of the mid-infrared radiation (of
order 0.1~Hz/min). We take this into account during spectroscopy.
The frequency instability of the mid-infrared radiation is less than
$2\times10^{-13}$ on time scales exceeding 1~s, and drops to below
$2\times10^{-14}$. The maser frequency itself is measured by comparison
with a 1~pulse-per-second signal obtained from GNSS satellites. The
maser frequency's deviation from 10~MHz is corrected for in the data
analysis.

\noindent 

\subsection{Experimental procedures}

The preparation and spectroscopy sequence is a variation of a previously
described procedure \cite{Alighanbari2020}.

A destructive spectroscopy is performed, where the vibrationally excited
molecular ions are subsequently dissociated by sequential excitation
by two additional lasers (resonance-enhanced multiphoton dissociation,
REMPD). The fractional decrease in number of trapped, intact HD\textsuperscript{+}
ions is determined. The first laser for REMPD is a continuous-wave
1475~nm laser tuned to the $(v'=1,\,N'=1)\rightarrow(v''=5,\,N''=2)$
transition, and the second one is a 266~nm continuous-wave laser
for subsequent dissociation.

Before the spectroscopy wave irradiation we perform 40~s of rotational
laser cooling to increase the population in the ground state $(v=0,N=0)$.

An important difference compared to our previous work is that the
spectroscopy radiation and the $1475\,$nm and 266~nm waves for REMPD
are not on simultaneously. Instead, we use a 5.2~s long sequence
during which spectroscopy and REMPD lasers are alternatingly blocked
and unblocked for 100~ms each. Power broadening of the transition
by the spectroscopy laser wave power is present but is kept moderate
(Fig. 2 in SM) by operating at sufficiently low power. However, performing
spectroscopy with high laser power is helpful for finding transitions.

\subsection{Vibrational transition}

The HD\textsuperscript{+} molecule harbors four angular momenta:
electron spin, proton spin, deuteron spin and rotational angular momentum.
The associated magnetic moments cause a hyperfine structure~{[}38{]}.
The ground rovibrational level has zero rotational angular momentum
$N=0$, giving rise to 4~spin states with total angular momentum
$F=0,\,1,\,2$ (this and other values are in part approximate quantum
numbers). The first excited vibrational level $(v'=1,\,N'=1)$ has
10 spin states, having values $F'=0,\,1,\,2,\,3$. Spin states are
$(2F+1)$-fold degenerate in zero magnetic field. A small applied
magnetic field leads to a quadratic Zeeman shift for $m_{{\rm F}}=0$
states, to a linear shift for the stretched states and to a combined
linear plus quadratic shift for the remaining Zeeman states.

We study two spin components having the same lower state (see Fig.~1
in SM): \\
line 12: $(v=0,N=0,G_{1}=1,G_{2}=2,F=2)\rightarrow(v'=1,N'=1,G_{1}'=1,G_{2}'=2,F'=1)$,\\
line 16: $(v=0,N=0,G_{1}=1,G_{2}=2,F=2)\rightarrow(v'=1,N'=1,G_{1}'=1,G_{2}'=2,F'=3)$.

Here $G_{1}$ refers to the sum of electron and proton spin, $G_{2}$
to the sum of $G_{1}$ and deuteron spin, and $F$ to the sum of $G_{2}$
and rotational angular momentum. We denote the transition frequencies
in zero external fields by $f_{{\rm 12}}$, $f_{{\rm 16}}$.

\subsection{\textit{Ab initio} theory of the HD\protect\textsuperscript{+} vibrational
transition}

An \emph{ab initio} transition frequency is composed of two contributions,
$f_{i}^{{\rm (theor)}}=f_{{\rm spin-avg}}^{{\rm (theor)}}+f_{{\rm spin},i}^{{\rm (theor)}}$.
The main one is the spin-averaged frequency $f_{{\rm spin-avg}}^{{\rm (theor)}}$, the difference between the level energies of the three-body system. The energies include the (essentially exact) non-relativistic energy plus relativistic, quantum-electrodynamic and finite-nuclear-size
corrections evaluated by perturbation theory~{[}39{]}.
The calculated value is  
\begin{align}
f_{{\rm spin-avg}}^{({\rm theor})} & =58\,605\,052\,163.9(5)_{{\rm theor,QED}}(13)_{{\rm CODATA2018}}\mbox{ kHz (case I)\,}.\label{eq:f_spin-avg}
\end{align}
CODATA 2018 values \cite{Tiesinga2019} of the fundamental constants
have been used (case I); the frequency value is updated compared to
the value presented in ref.~{[}40{]}. The contributions
of relative order $(\alpha^{0},\alpha^{2},\alpha^{3},\alpha^{4},\alpha^{5},\alpha^{6})$
to eq.~(\ref{eq:f_spin-avg}) are $(58\,604\,301\,249.69,$ $1\,003\,554.55,$
$-250\,978.39,$ $-1\,770.95,$ $109.52$, $-0.77)$~kHz. The proton
size contributes $-17.17(8)_{{\rm CODATA2018}}\,{\rm kHz}$, the
deuteron size $-109.67(8)_{{\rm CODATA2018}}\,{\rm kHz}$; these
contributions are included in the term of relative order $\alpha^{2}$.
Several further corrections were added, e.g. stemming from the polarizability
of the deuteron, from second-order vibrational contributions (of relative
order $\alpha^{8}$), etc., amounting to $0.26\,{\rm kHz}$. The first
uncertainty indicated in eq.~(\ref{eq:f_spin-avg}), $u(f_{{\rm spin-avg}}^{({\rm theor})})=0.5\,$kHz
($8\times10^{-12}$ fractionally), is an estimate of the unevaluated
QED contributions. The second uncertainty is due to the uncertainties
of the fundamental constants. Here, the most important contribution
is from the uncertainty of $m_{{\rm p}}/m_{{\rm e}}$, 1.1~kHz.

To obtain the functional dependence of the spin-averaged frequency
on $\mu_{{\rm p}}=m_{{\rm p}}/m_{{\rm e}}$ we calculated \emph{ab
initio} the derivative $\partial{\rm ln}f_{{\rm spin-avg}}^{{\rm (theor)}}(\mu_{{\rm p}})/\partial{\rm ln}\mu_{{\rm p}}|_{m_{{\rm d}}/m_{{\rm p}}=const.}=-0.4846$;
it is close to the Born-Oppenheimer value $-\frac{1}{2}$.

The hyperfine structure in both the ground and the excited vibrational
level results in the hyperfine shifts $f_{{\rm spin},i}$. The main
contribution to the hyperfine structure of each level comes from the
Fermi contact interaction between electron spin and proton spin, followed
by the contact interaction between the electron spin and the deuteron
spin. These and further interactions are described by an effective
Hamiltonian~{[}38{]} and quantified by corresponding spin
structure coefficients: ${\cal E}_{4},{\cal E}_{5}$ for the $(v=0,N=0)$
level and ${\cal E}_{1}',\dots{\cal E}_{9}'$ for the ($v'=1,N'=1$)
level. The different strengths of the interactions in the two levels
lead to a nonzero hyperfine shift. Given the aim of this work, it
is important to have an accurate prediction for the hyperfine shift,
implying the need for an accurate \textit{ab initio} computation of
the coefficients ${\cal E}_{i}$. A high-precision calculation~{[}41{]}
provides the set of Fermi contact interaction coefficients $S_{a}$:
$({\cal E}_{4},{\cal E}_{4}',{\cal E}_{5},{\cal E}_{5}')$. The values
${\cal E}_{4},{\cal E}_{5}$ for the ground state $(v=0,N=0)$ have
already played a role in rotational spectroscopy \cite{Alighanbari2020}.
For the level $(v'=1,N'=1)$, ${\cal E}_{4}',{\cal E}_{5}'$ are computed
here for the first time and are given in SM. These coefficients have
fractional theoretical uncertainties, $\epsilon_{{\rm F}}$, of order
$\alpha^{3}$. We estimate $\epsilon_{{\rm F}}=1\times10^{-6}$. The
set $S_{a}$ contributes approximately $\delta_{S_{a},i}=\epsilon_{{\rm F}}(\gamma_{i,4}{\cal E}_{4},\gamma_{i,4}'{\cal E}_{4}',\gamma_{i,5}{\cal E}_{5},\gamma_{i,5}'{\cal E}_{5}')\simeq$$(0.23,0.23,0.07,0.07)$~kHz
to both $f_{{\rm spin},12}^{({\rm theor})}$ and $f_{{\rm spin},16}^{({\rm theor})}$.
The sensitivities $\gamma_{i,k},\gamma_{i,k}'$ are taken from Tab.~1
in SM.

The coefficient ${\cal E}_{1}'$ is the next-largest spin coefficient.
It has recently been the subject of intense theory work beyond the
Breit-Pauli approximation, resulting in the value shown in Tab.~1
of SM, with an improved theory uncertainty of $u_{1}'=0.05$~kHz~{[}29, 42{]}. The set $S_{b}$
comprising the remaining spin coefficients ${\cal E}_{2}'$, ${\cal E}_{3}',\,{\cal E}_{6}'$,
${\cal E}_{7}',\,{\cal E}_{8}',\,{\cal E}_{9}'$, has been computed
within the Breit-Pauli approximation. It neglects terms of relative
order $\alpha^{2}$; therefore we assume that the fractional uncertainties
of the set's elements are $\epsilon_{0}=\alpha^{2}$. As a result,
the dominant contribution of the set $({\cal E}_{1}',\,S_{b})$ to
the uncertainties of $f_{{\rm spin},12}^{({\rm theor})}$ and $f_{{\rm spin},16}^{({\rm theor})}$ is from the coefficient ${\cal E}_{7}'$ and ${\cal E}_{6}'$,
respectively.

In summary, the individual spin frequencies are  
\begin{align}
f_{{\rm spin,12}}^{{\rm (theor)}} & =-38\,686.1(8)_{{\rm theor,spin}}\,{\rm kHz}\,,\nonumber \\
f_{{\rm spin,16}}^{{\rm (theor)}} & =\phantom{-3}2\,607.7(9)_{{\rm theor,spin}}\,{\rm kHz}\,.\label{eq:f_spin-1}
\end{align}
The uncertainties were here estimated as
\begin{align*}
u_{{\rm theor,spin}}(f_{{\rm spin,}i}^{({\rm theor})}) & =|\gamma_{i,1}'u_{1}'|+\epsilon_{\text{0}}\sum_{k=2,3,6,7,8,9}|\gamma_{i,k}'{\cal E}_{k}'|+\epsilon_{{\rm F}}\sum_{k=4,5}|\gamma_{i,k}'{\cal E}_{k}'|+|\gamma_{i,k}{\cal E}_{k}|\,.
\end{align*}
The uncertainties from the various spin coefficients are not added
quadratically because we cannot assume that they are independent random
errors. The sum of eq.~(\ref{eq:f_spin-avg}) and eq.~(\ref{eq:f_spin-1})
is eq.~(\ref{eq:f_theor_total}).

The hyperfine splitting $f_{{\rm spin},16}^{({\rm theor})}-f_{{\rm spin},12}^{({\rm theor})}$
depends only on the spin coefficients of the upper level, since line~12
and line~16 have a common lower state. The uncertainty of the splitting
is not affected by the uncertainties of ${\cal E}_{4}'$, ${\cal E}_{5}'$
since $\gamma_{12,4}'=\gamma_{16,4}'$, $\gamma_{12,5}'\approx\gamma_{16,5}'$.
The uncertainty is mostly determined by the uncertainties of the next-next-largest
coefficients ${\cal E}_{6}'$ and ${\cal E}_{7}'$, 
\begin{align}
f_{{\rm spin},16}^{({\rm theor})}-f_{{\rm spin},12}^{({\rm theor})} & =41\,293.81(44)_{{\rm theor,spin}}\,{\rm kHz}\,.\label{eq:theory fspin16 - fspin12}
\end{align}
The difference of the two measured vibrational frequencies is 
\[
f_{{\rm spin},16}^{({\rm exp})}-f_{{\rm spin},12}^{({\rm exp})}=41\,294.06(32)_{{\rm exp}}\,{\rm kHz},
\]
in very good agreement with the prediction.

\subsection{Systematic shifts}

We have previously computed  several external-field shifts of vibrational
transition frequencies: Zeeman shift~{[}43{]}, electric
quadrupole shift~{[}44{]}, d.c.~Stark shift, black-body
radiation shift, and spin-state dependence of the d.c.~Stark and
light shift~{[}45{]}. Black-body radiation shift and
electric quadrupole shift are negligible.

The Zeeman shifts for the components $m_{F}=0\rightarrow m_{F}'=0$
of line~12 and 16 are $-2.9\,$kHz/G$^{2}$ and $-117\,$kHz/G$^{2}$,
respectively. For the transitions between stretched states, $m_{F}=\pm2\rightarrow m_{F}'=\pm3$
of line~16, the Zeeman shift is purely linear, with coefficients
$\mp0.55\,$kHz/G.

The light shift of the lower and upper vibrational levels due to the
313~nm cooling radiation can be computed using the \emph{ab initio}
frequency-dependent polarisability values. The latter can be calculated
using the procedure described in~{[}45{]}. For the lower
level, they have already been reported in ref.~\cite{Alighanbari2020}.
For the upper level the scalar ($s$) and tensor ($t$) polarisabilities
are $\alpha_{s}(v'=1,\,N'=1,\,\lambda=313\,{\rm nm})=4.475$ atomic
units, $\alpha_{t}(v'=1,\,N'=1,\,\lambda=313\,{\rm nm})=-1.442$ atomic
units. They are within 1.5 atomic units of the values for the lower
level. The calculated light shift is therefore negligible.

Following ref.~\cite{Alighanbari2020}, we also investigated the
effect on the transition frequency of a substantial displacement of
the atomic ion cluster in radial direction. When doing so, the HD\textsuperscript{+}
ion string does not visibly shift. The deformation was obtained by
changing substantially the d.c.~voltages on the two lower trap electrodes
by 5~V, shifting the beryllium cluster by approximately 0.10~mm
orthogonally to the trap axis. No frequency shift was observed, with
an upper bound of \textbf{$0.4\,{\rm kHz}$} (Fig.~3 in SM). The
upper bound of the corresponding frequency shift of the \emph{rotational}
transition \cite{Alighanbari2020} was significantly smaller than
this value. We expect the shifts to be of similar absolute magnitude
for the two cases. We therefore do not apply any correction or uncertainty
for this perturbation in our analysis. Thus, the only systematic shifts
corrected for in this work are Zeeman and trap-induced d.c. Stark
shifts, see main text.

For every measured transition line we assigned one-half of the fitted
full width as statistical uncertainty of the line center transition
frequency.

\subsection{Composite frequency}

The spin theory uncertainty of the composite frequency is computed
as
\begin{align*}
u_{{\rm theor,spin}}(f_{{\rm spin-avg}}^{({\rm exp})}) & =|\sum_{i=12,16}b_{i}\gamma_{i,1}'u_{1}'|+\epsilon_{\text{0}}\sum_{k=2,3,6,7,8,9}|\sum_{i=12,16}b_{i}\gamma_{i,k}'{\cal E}_{k}'|\\
 & \phantom{=}+\epsilon_{{\rm F}}\sum_{k=4,5}(\,|\sum_{i=12,16}b_{i}\gamma_{i,k}'{\cal E}_{k}'|+|\sum_{i=12,16}b_{i}\gamma_{i,k}{\cal E}_{k}|\,)\,.
\end{align*}
where $b_{16}=1-b_{12}$.

\subsection{Previous results on high-resolution spectroscopy and mid-infrared
spectroscopy sources}

We briefly review the work achieving the highest line resolutions
without Lamb-Dicke or resolved-carrier regime. For gas-phase neutral
molecules at non-cryogenic temperature, line resolutions  up to 
  $9\times10^{10}$ were achieved for vibrational transitions~{[}46-48{]}
and  $7\times10^{9}$ for electronic transitions~{[}49{]},
using saturation or Ramsey spectroscopy (for a special case, see~{[}50{]}). In contrast, for atomic and molecular ions under
gaseous conditions the best  resolutions have been significantly
lower. For example, saturation spectroscopy of gas-phase molecular
ions has led to $2\times10^{6}$ resolution (absolute full linewidths
at the 50~MHz level)~{[}51{]}. Using the ion beam technique,
electronic and vibrational transitions have been observed with resolutions
up to $2.5\times10^{6}$~{[}52{]} and  $4\times10^{6}$~{[}53, 54{]}, respectively. In ion traps equipped with
cooling by collisions with cold helium buffer gas, $3\times10^{6}$
resolution in vibrational spectroscopy (30~MHz linewidth) has been
reached~{[}55{]}.

Previously, one-photon vibrational spectroscopy of sympathetically
cooled molecular ion ensembles has been reported~{[}30, 33, 56-59{]}.
However, in these experiments, the spectroscopy wave was irradiated
along the trap axis and/or the laser linewidth was high, leading to
highest line resolution of $2\times10^{7}$ (3~MHz absolute linewidth)
\cite{Bressel2012}.

Difference-frequency sources emitting at shorter wavelength were reported
earlier (see ref.~{[}60{]} for an overview), but had larger
linewidth and/or higher frequency instability than the present one.
At 3.4~$\mu$m, linewidths of 60~kHz~{[}61{]} and more
recently of 3.5~kHz~{[}62{]} were achieved. An alternative
approach are quantum cascade lasers. They have been shown to allow
ultra-narrow linewidth and precise frequency calibration~{[}63, 64{]}
(see ref.~{[}65{]} for a review).

\end{NoHyper}
\smallskip{}

\begin{acknowledgments}
We are indebted to E. Wiens for assistance with the frequency comb
measurements. We are very grateful to J.-P. Karr for communicating
the value of ${\cal E}_{1}'$ before publication. This work has received
funding from the European Research Council (ERC) under the European
Union\textquoteright s Horizon 2020 research and innovation programme
(grant agreement No. 786306, ``PREMOL''), from the Deutsche Forschungsgemeinschaft
(Schi 431/23-1) and from both DFG and the state of Nordrhein-Westfalen
via grant INST-208/737-1 FUGG. I.K. was partly supported by FP7-2013-ITN
\textquotedblleft COMIQ\textquotedblright{} (Grant No. 607491). V.I.K.
acknowledges support from the Russian Foundation for Basic Research
under Grant No. 19-02-00058-a.
\end{acknowledgments}

\textbf{\smallskip{}
Author Contributions}

I.K., M.G.H., and S.S. developed the laser system, I.K., S.A., and
G.S.G. performed the experiments and analyzed the data, I.K., S.S.
and V.I.K. performed theoretical calculations, S.S. conceived the
study, supervised the work and wrote the paper. All authors contributed
to editing of the manuscript.

\textbf{\smallskip{}
Data Availability}\\
Source data are available for this paper. All other data that support
the plots within this paper and other findings of this study are
available from the corresponding author upon reasonable request.

\textbf{\smallskip{}
Additional Information}

Supplementary Information is available for the paper at [DOI link to be inserted]

Correspondence and requests for materials should be addressed to S.~Schiller.

The authors declare no competing financial or non-financial interests.

Reprints and permissions information is available at www.nature.com/reprints.

\emph{\medskip{}
}

\textbf{References for Methods}

\noindent
{[}36{]}
\bibinfo{author}{Chen, Q.-F.} \emph{et~al.}
\newblock \bibinfo{title}{A compact, robust, and transportable ultra-stable
  laser with a fractional frequency instability of $1\times10^{-15}$}.
\newblock \emph{\bibinfo{journal}{Rev. Sci. Instrum.}}
  \textbf{\bibinfo{volume}{85}}, \bibinfo{pages}{113107}
  (\bibinfo{year}{2014}).

\noindent
{[}37{]}
\bibinfo{author}{Wiens, E.}, \bibinfo{author}{Nevsky, A.~Y.} \&
  \bibinfo{author}{Schiller, S.}
\newblock \bibinfo{title}{Resonator with ultrahigh length stability as a probe
  for equivalence-principle-violating physics}.
\newblock \emph{\bibinfo{journal}{Phys. Rev. Lett.}}
  \textbf{\bibinfo{volume}{117}}, \bibinfo{pages}{271102}
  (\bibinfo{year}{2016}).

\noindent
{[}38{]}
\bibinfo{author}{Bakalov, D.}, \bibinfo{author}{Korobov, V.~I.} \&
  \bibinfo{author}{Schiller, S.}
\newblock \bibinfo{title}{High-precision calculation of the hyperfine structure
  of the {HD}$^+$ ion}.
\newblock \emph{\bibinfo{journal}{Phys. Rev. Lett.}}
  \textbf{\bibinfo{volume}{97}}, \bibinfo{pages}{243001}
  (\bibinfo{year}{2006}).

\noindent
{[}39{]}
\bibinfo{author}{Korobov, V.~I.}, \bibinfo{author}{Hilico, L.} \&
  \bibinfo{author}{Karr, J.-P.}
\newblock \bibinfo{title}{{Fundamental Transitions and Ionization Energies of
  the Hydrogen Molecular Ions with Few ppt Uncertainty}}.
\newblock \emph{\bibinfo{journal}{Phys. Rev. Lett.}}
  \textbf{\bibinfo{volume}{118}}, \bibinfo{pages}{233001}
  (\bibinfo{year}{2017}).

\noindent
{[}40{]}
\bibinfo{author}{Aznabayev, D.~T.}, \bibinfo{author}{Bekbaev, A.~K.} \&
  \bibinfo{author}{Korobov, V.~I.}
\newblock \bibinfo{title}{{Leading-order relativistic corrections to the
  rovibrational spectrum of ${\text{H}}_{2}{\phantom{\rule{0.16em}{0ex}}}^{+}$
  and ${\mathrm{HD}}^{+}$ molecular ions}}.
\newblock \emph{\bibinfo{journal}{Phys. Rev. A}} \textbf{\bibinfo{volume}{99}},
  \bibinfo{pages}{012501} (\bibinfo{year}{2019}).

\noindent
{[}41{]}
\bibinfo{author}{Korobov, V.~I.}, \bibinfo{author}{Koelemeij, J. C.~J.},
  \bibinfo{author}{Hilico, L.} \& \bibinfo{author}{Karr, J.-P.}
\newblock \bibinfo{title}{{Theoretical Hyperfine Structure of the Molecular
  Hydrogen Ion at the 1 ppm Level}}.
\newblock \emph{\bibinfo{journal}{Phys. Rev. Lett.}}
  \textbf{\bibinfo{volume}{116}}, \bibinfo{pages}{053003}
  (\bibinfo{year}{2016}).

\noindent
{[}42{]}
\bibinfo{author}{Karr, J.-P.} \& \bibinfo{author}{Haidar, M.}
\newblock \bibinfo{title}{private comm.} (\bibinfo{year}{2020}).

\noindent
{[}43{]}
\bibinfo{author}{Bakalov, D.}, \bibinfo{author}{Korobov, V.} \&
  \bibinfo{author}{Schiller, S.}
\newblock \bibinfo{title}{Magnetic field effects in the transitions of the
  {HD}$^+$ molecular ion and precision spectroscopy}.
\newblock \emph{\bibinfo{journal}{J. Phys. B: At. Mol. Opt. Phys.}}
  \textbf{\bibinfo{volume}{44}}, \bibinfo{pages}{025003}
  (\bibinfo{year}{2011}).
\newblock \bibinfo{note}{Corrigendum: {\it J. Phys. B: At. Mol. Opt. Phys.}
  {\bf 45}, 049501 (2012)}.

\noindent
{[}44{]}
\bibinfo{author}{Bakalov, D.} \& \bibinfo{author}{Schiller, S.}
\newblock \bibinfo{title}{The electric quadrupole moment of molecular hydrogen
  ions and their potential for a molecular ion clock}.
\newblock \emph{\bibinfo{journal}{Appl. Phys. B}}
  \textbf{\bibinfo{volume}{114}}, \bibinfo{pages}{213--230}
  (\bibinfo{year}{2014}).

\noindent
{[}45{]}
\bibinfo{author}{Schiller, S.}, \bibinfo{author}{Bakalov, D.},
  \bibinfo{author}{Bekbaev, A.~K.} \& \bibinfo{author}{Korobov, V.~I.}
\newblock \bibinfo{title}{Static and dynamic polarizability and the {S}tark and
  blackbody-radiation frequency shifts of the molecular hydrogen ions {H$_2^+$,
  HD$^+$, and D$_2^+$}}.
\newblock \emph{\bibinfo{journal}{Phys. Rev. A}} \textbf{\bibinfo{volume}{89}},
  \bibinfo{pages}{052521} (\bibinfo{year}{2014}).

\noindent
{[}46{]}
\bibinfo{author}{Hall, J.~L.}, \bibinfo{author}{Bord{\'e}, C.~J.} \&
  \bibinfo{author}{Uehara, K.}
\newblock \bibinfo{title}{Direct optical resolution of the recoil effect using
  saturated absorption spectroscopy}.
\newblock \emph{\bibinfo{journal}{Phys. Rev. Lett.}}
  \textbf{\bibinfo{volume}{37}}, \bibinfo{pages}{1339--1342}
  (\bibinfo{year}{1976}).

\noindent
{[}47{]}
\bibinfo{author}{Salomon, C.}, \bibinfo{author}{Br{\'e}ant, C.},
  \bibinfo{author}{Bord{\'e}, C.} \& \bibinfo{author}{Barger, R.}
\newblock \bibinfo{title}{Ramsey fringes using transitions in the visible and
  10-$\mu$-m spectral regions - experimental methods}.
\newblock \emph{\bibinfo{journal}{Journal de Physique Colloques}}
  \textbf{\bibinfo{volume}{{42}}}, \bibinfo{pages}{{3--14}}
  (\bibinfo{year}{1981}).

\noindent
{[}48{]}
\bibinfo{author}{Bagayev, S.~N.}, \bibinfo{author}{Baklanov, A.~E.},
  \bibinfo{author}{Chebotayev, V.~P.} \& \bibinfo{author}{Dychkov, A.~S.}
\newblock \bibinfo{title}{Superhigh resolution spectroscopy in methane with
  cold molecules}.
\newblock \emph{\bibinfo{journal}{Appl. Phys. B}}
  \textbf{\bibinfo{volume}{48}}, \bibinfo{pages}{31--35}
  (\bibinfo{year}{1989}).

\noindent
{[}49{]}
\bibinfo{author}{Cheng, W.-Y.}, \bibinfo{author}{Chen, L.},
  \bibinfo{author}{Yoon, T.~H.}, \bibinfo{author}{Hall, J.~L.} \&
  \bibinfo{author}{Ye, J.}
\newblock \bibinfo{title}{{Sub-Doppler molecular-iodine transitions near the
  dissociation limit (523 - 498 nm)}}.
\newblock \emph{\bibinfo{journal}{Opt. Lett.}} \textbf{\bibinfo{volume}{27}},
  \bibinfo{pages}{571--573} (\bibinfo{year}{2002}).

\noindent
{[}50{]}
\bibinfo{author}{Bagayev, S.~N.} \emph{et~al.}
\newblock \bibinfo{title}{Second-order doppler-free spectroscopy}.
\newblock \emph{\bibinfo{journal}{Appl. Phys. B}}
  \textbf{\bibinfo{volume}{52}}, \bibinfo{pages}{63--66}
  (\bibinfo{year}{1991}).

\noindent
{[}51{]}
\bibinfo{author}{Markus, C.~R.}, \bibinfo{author}{Kocheril, P.~A.} \&
  \bibinfo{author}{McCall, B.~J.}
\newblock \bibinfo{title}{{Sub-Doppler rovibrational spectroscopy of the
  $\nu_1$ fundamental band of {D$_2$H$^+$}}}.
\newblock \emph{\bibinfo{journal}{J. Mol. Spectrosc.}}
  \textbf{\bibinfo{volume}{355}}, \bibinfo{pages}{8--13}
  (\bibinfo{year}{2019}).

\noindent
{[}52{]}
\bibinfo{author}{Mills, A.~A.} \emph{et~al.}
\newblock \bibinfo{title}{Ultra-sensitive high-precision spectroscopy of a fast
  molecular ion beam}.
\newblock \emph{\bibinfo{journal}{J. Chem. Phys.}}
  \textbf{\bibinfo{volume}{135}}, \bibinfo{pages}{224201}
  (\bibinfo{year}{2011}).

\noindent
{[}53{]}
\bibinfo{author}{Wing, W.~H.}, \bibinfo{author}{Ruff, G.~A.},
  \bibinfo{author}{Lamb, W.~E.} \& \bibinfo{author}{Spezeski, J.~J.}
\newblock \bibinfo{title}{{Observation of the Infrared Spectrum of the Hydrogen
  Molecular Ion HD$^{+}$}}.
\newblock \emph{\bibinfo{journal}{Phys. Rev. Lett.}}
  \textbf{\bibinfo{volume}{36}}, \bibinfo{pages}{1488--1491}
  (\bibinfo{year}{1976}).

\noindent
{[}54{]}
\bibinfo{author}{Coe, J.~V.} \emph{et~al.}
\newblock \bibinfo{title}{{Sub-Doppler direct infrared laser absorption
  spectroscopy in fast ion beams: The fluorine hyperfine structure of
  {HF$^+$}}}.
\newblock \emph{\bibinfo{journal}{J. Chem. Phys.}}
  \textbf{\bibinfo{volume}{90}}, \bibinfo{pages}{3893--3902}
  (\bibinfo{year}{1989}).

\noindent
{[}55{]}
\bibinfo{author}{Markus, C.~R.}, \bibinfo{author}{Thorwirth, S.},
  \bibinfo{author}{Asvany, O.} \& \bibinfo{author}{Schlemmer, S.}
\newblock \bibinfo{title}{{High-resolution double resonance action spectroscopy
  in ion traps: vibrational and rotational fingerprints of {CH$_2$NH$_2^+$}}}.
\newblock \emph{\bibinfo{journal}{Phys. Chem. Chem. Phys.}}
  \textbf{\bibinfo{volume}{21}}, \bibinfo{pages}{26406--26412}
  (\bibinfo{year}{2019}).

\noindent
{[}56{]}
\bibinfo{author}{Roth, B.}, \bibinfo{author}{Koelemeij, J. C.~J.},
  \bibinfo{author}{Daerr, H.} \& \bibinfo{author}{Schiller, S.}
\newblock \bibinfo{title}{Rovibrational spectroscopy of trapped molecular
  hydrogen ions at millikelvin temperatures}.
\newblock \emph{\bibinfo{journal}{Phys. Rev. A}} \textbf{\bibinfo{volume}{74}},
  \bibinfo{pages}{040501} (\bibinfo{year}{2006}).

\noindent
{[}57{]}
\bibinfo{author}{Koelemeij, J. C.~J.}, \bibinfo{author}{Noom, D. W.~E.},
  \bibinfo{author}{de~Jong, D.}, \bibinfo{author}{Haddad, M.~A.} \&
  \bibinfo{author}{Ubachs, W.}
\newblock \bibinfo{title}{{Observation of the $v' = 8\leftarrow v = 0$
  vibrational overtone in cold trapped HD$^+$}}.
\newblock \emph{\bibinfo{journal}{Appl. Phys. B}}
  \textbf{\bibinfo{volume}{107}}, \bibinfo{pages}{1075--1085}
  (\bibinfo{year}{2012}).

\noindent
{[}58{]}
\bibinfo{author}{Biesheuvel, J.} \emph{et~al.}
\newblock \bibinfo{title}{{Probing QED and fundamental constants through laser
  spectroscopy of vibrational transitions in {HD}$^+$}}.
\newblock \emph{\bibinfo{journal}{Nat. Commun.}} \textbf{\bibinfo{volume}{7}},
  \bibinfo{pages}{10385} (\bibinfo{year}{2016}).

\noindent
{[}59{]}
\bibinfo{author}{Calvin, A.~T.} \emph{et~al.}
\newblock \bibinfo{title}{{Rovibronic Spectroscopy of Sympathetically Cooled
  {$^{40}$CaH$^+$}}}.
\newblock \emph{\bibinfo{journal}{J. Phys. Chem. A}}
  \textbf{\bibinfo{volume}{122}}, \bibinfo{pages}{3177--3181}
  (\bibinfo{year}{2018}).

\noindent
{[}60{]}
\bibinfo{author}{Liao, C.-C.}, \bibinfo{author}{Lien, Y.-H.},
  \bibinfo{author}{Wu, K.-Y.}, \bibinfo{author}{Lin, Y.-R.} \&
  \bibinfo{author}{Shy, J.-T.}
\newblock \bibinfo{title}{Widely tunable difference frequency generation source
  for high-precision mid-infrared spectroscopy}.
\newblock \emph{\bibinfo{journal}{Opt. Express}} \textbf{\bibinfo{volume}{21}},
  \bibinfo{pages}{9238--9246} (\bibinfo{year}{2013}).

\noindent
{[}61{]}
\bibinfo{author}{Takahata, K.} \emph{et~al.}
\newblock \bibinfo{title}{Absolute frequency measurement of sub-doppler
  molecular lines using a 3.4 $\mu$m difference-frequency-generation
  spectrometer and a fiber-based frequency comb}.
\newblock \emph{\bibinfo{journal}{Phys. Rev. A}} \textbf{\bibinfo{volume}{80}},
  \bibinfo{pages}{032518} (\bibinfo{year}{2009}).

\noindent
{[}62{]}
\bibinfo{author}{Sera, H.} \emph{et~al.}
\newblock \bibinfo{title}{{Sub-Doppler resolution mid-infrared spectroscopy
  using a difference-frequency-generation source spectrally narrowed by laser
  linewidth transfer}}.
\newblock \emph{\bibinfo{journal}{Opt. Lett.}} \textbf{\bibinfo{volume}{40}},
  \bibinfo{pages}{5467--5470} (\bibinfo{year}{2015}).

\noindent
{[}63{]}
\bibinfo{author}{Hansen, M.~G.}, \bibinfo{author}{Magoulakis, E.},
  \bibinfo{author}{Chen, Q.-F.}, \bibinfo{author}{Ernsting, I.} \&
  \bibinfo{author}{Schiller, S.}
\newblock \bibinfo{title}{Quantum cascade laser-based mid-{IR} frequency
  metrology system with ultra-narrow linewidth and $1\times 10^{-13}$-level
  frequency instability}.
\newblock \emph{\bibinfo{journal}{Opt. Lett.}} \textbf{\bibinfo{volume}{40}},
  \bibinfo{pages}{2289--2292} (\bibinfo{year}{2015}).

\noindent
{[}64{]}
\bibinfo{author}{Argence, B.} \emph{et~al.}
\newblock \bibinfo{title}{{Quantum cascade laser frequency stabilization at the
  sub-Hz level}}.
\newblock \emph{\bibinfo{journal}{Nat. Photonics}}
  \textbf{\bibinfo{volume}{9}}, \bibinfo{pages}{456--460}
  (\bibinfo{year}{2015}).

\noindent
{[}65{]}
\bibinfo{author}{Borri, S.} \emph{et~al.}
\newblock \bibinfo{title}{High-precision molecular spectroscopy in the
  mid-infrared using quantum cascade lasers}.
\newblock \emph{\bibinfo{journal}{Appl. Phys. B}}
  \textbf{\bibinfo{volume}{125}}, \bibinfo{pages}{18} (\bibinfo{year}{2019}).

\end{document}


\title{\textcolor{black}{Supplementary Information for ``Proton-electron mass ratio by high-resolution optical spectroscopy of ion ensembles in the resolved-carrier regime"}}
\author{I. Kortunov}
\affiliation{Institut für Experimentalphysik, Heinrich-Heine-Universität Düsseldorf,
40225 Düsseldorf, Germany}
\author{S. Alighanbari}
\affiliation{Institut für Experimentalphysik, Heinrich-Heine-Universität Düsseldorf,
40225 Düsseldorf, Germany}
\author{M.~G. Hansen}
\affiliation{Institut für Experimentalphysik, Heinrich-Heine-Universität Düsseldorf,
40225 Düsseldorf, Germany}
\author{G.~S.~Giri}
\affiliation{Institut für Experimentalphysik, Heinrich-Heine-Universität Düsseldorf,
40225 Düsseldorf, Germany}
\author{V.~I. Korobov}
\affiliation{Bogoliubov Laboratory of Theoretical Physics, Joint Institute for
Nuclear Research, 141980 Dubna, Russia}
\author{S. Schiller}
\email[Corresponding Author, e-mail:~]{step.schiller@hhu.de}
\affiliation{Institut für Experimentalphysik, Heinrich-Heine-Universität Düsseldorf,
40225 Düsseldorf, Germany}
\maketitle

\begin{NoHyper}
\begin{center}
\includegraphics[scale=0.65]{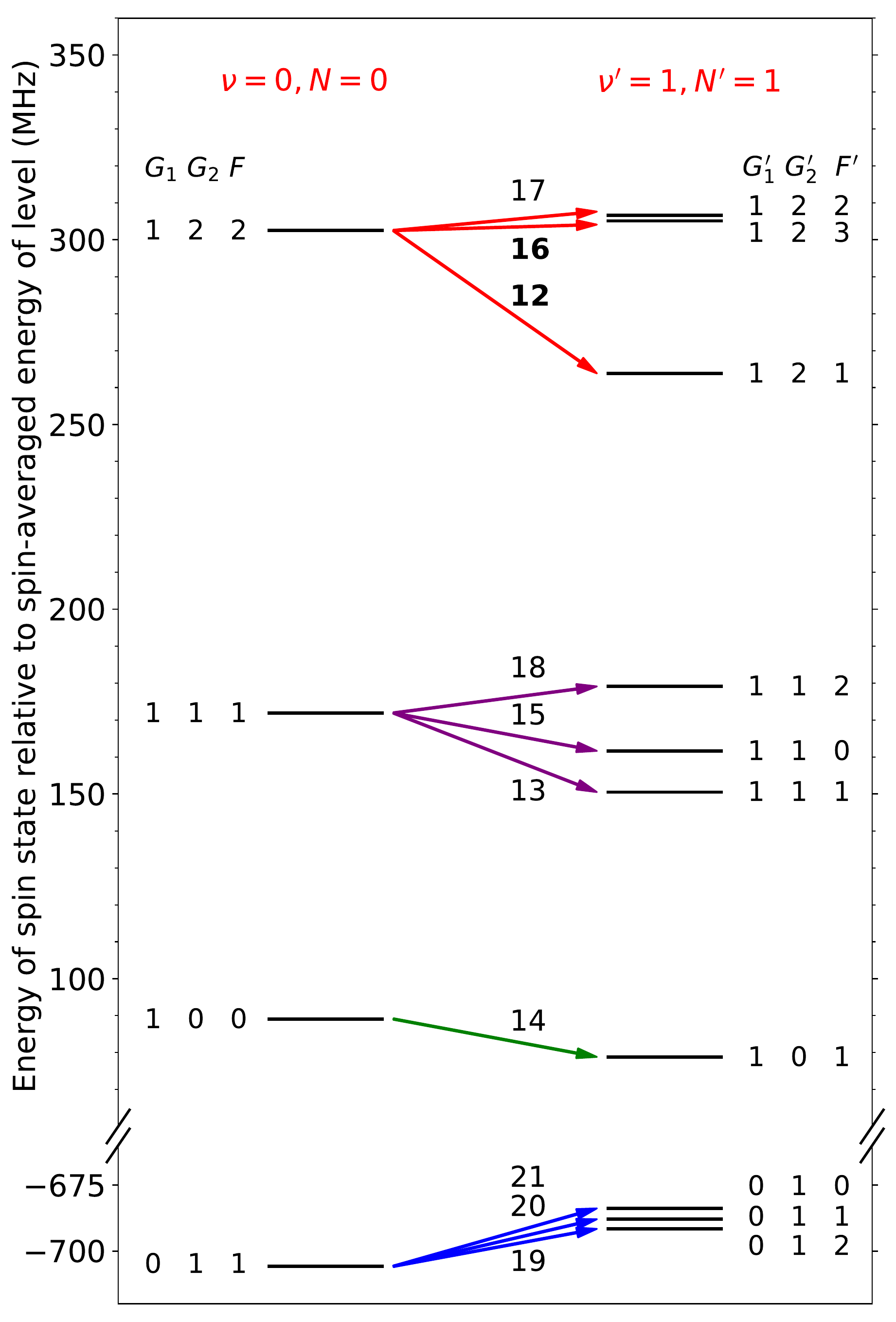}
\par\end{center}

\noindent Supplementary Figure 1. \textbf{Hyperfine structure energy levels of HD}\textsuperscript{\textbf{+}}\textbf{
in the two relevant levels.} Shown are the ground vibrational level
$(v=0,N=0)$ (left) and the first excited vibrational level $(v'=1,N'=1)$
(right) of the $^{2}\Sigma_{g}^{+}$ electronic state. The numbers
next to the arrows indicate the line numbers. In this work, lines~12
and 16 were detected and measured.\textbf{\clearpage\pagebreak}

{
\centering
\scriptsize
\begin{tabular*}{1.01\textwidth}{|c @{\extracolsep{\fill}}|c|c|c|c|c|c|c|c|c|c|c|c|}

\hline 
 &  & ${\cal E}_{1}'$ & ${\cal E}_{2}'$ & ${\cal E}_{3}'$ & ${\cal E}_{4}'$ & ${\cal E}_{5}'$ & ${\cal E}_{6}'$ & ${\cal E}_{7}'$ & ${\cal E}_{8}'$ & ${\cal E}_{9}'$ & ${\cal E}_{4}$ & ${\cal E}_{5}$ \\ 
 
 &  & 30.28083 & -0.03046 & -0.004664 & 903.36802 & 138.91049 & 8.13669 & 1.24894 & -0.002945 & 0.005659 & 925.39588 & 142.28781 \\ 
\hline 
Line $i$ & $f_{\mathrm{spin},i}^{\mathrm{(theor)}}$ & $\gamma_{i,1}'$ & $\gamma_{i,2}'$ & $\gamma_{i,3}'$ & $\gamma_{i,4}'$ & $\gamma_{i,5}'$ & $\gamma_{i,6}'$ & $\gamma_{i,7}'$ & $\gamma_{i,8}'$ & $\gamma_{i,9}'$ & $\gamma_{i,4}$ & $\gamma_{i,5}$ \\ 
\hline 
12 & -38.68609 & -0.575 & -0.565 & -1.715 & 0.250 & 0.430 & -0.011 & -3.369 & -3.306 & -2.909 & 0.250 & 0.500 \\ 
 
16 & 2.60772 & 0.500 & 0.500 & 1.000 & 0.250 & 0.500 & -0.500 & -1.000 & -1.000 & -0.500 & 0.250 & 0.500 \\ 
\hline 
\end{tabular*} 
}

\noindent Supplementary Table 1. \textbf{Spin hamiltonian coefficients, spin structure frequencies,
and spin frequency derivatives.} ${\cal E}_{k}'$ are the coefficients
of the spin Hamiltonian~{[}45{]} for the $(v'=1,N'=1)$
level, in MHz. ${\cal E}_{k}$ are the coefficients for the ro-vibrational
ground state $(v=0,N=0)$, already reported in~{[}13{]}.
$f_{{\rm spin}}^{({\rm theor})}$ is the theoretical spin structure
frequency in MHz. $\gamma$ are the dimensionless sensitivities of
the spin structure frequencies  to the various spin Hamiltonian coefficients.
 $\gamma_{i,k}'=\partial f_{{\rm spin,}i}^{({\rm theor})}/\partial{\cal E}_{k}'$
refers to the upper state, $\gamma_{i,k}=-\partial f_{{\rm spin,}i}^{({\rm theor})}/\partial{\cal E}_{k}$
to the lower state.

\clearpage

\begin{center}
\textbf{\includegraphics[width=\columnwidth]{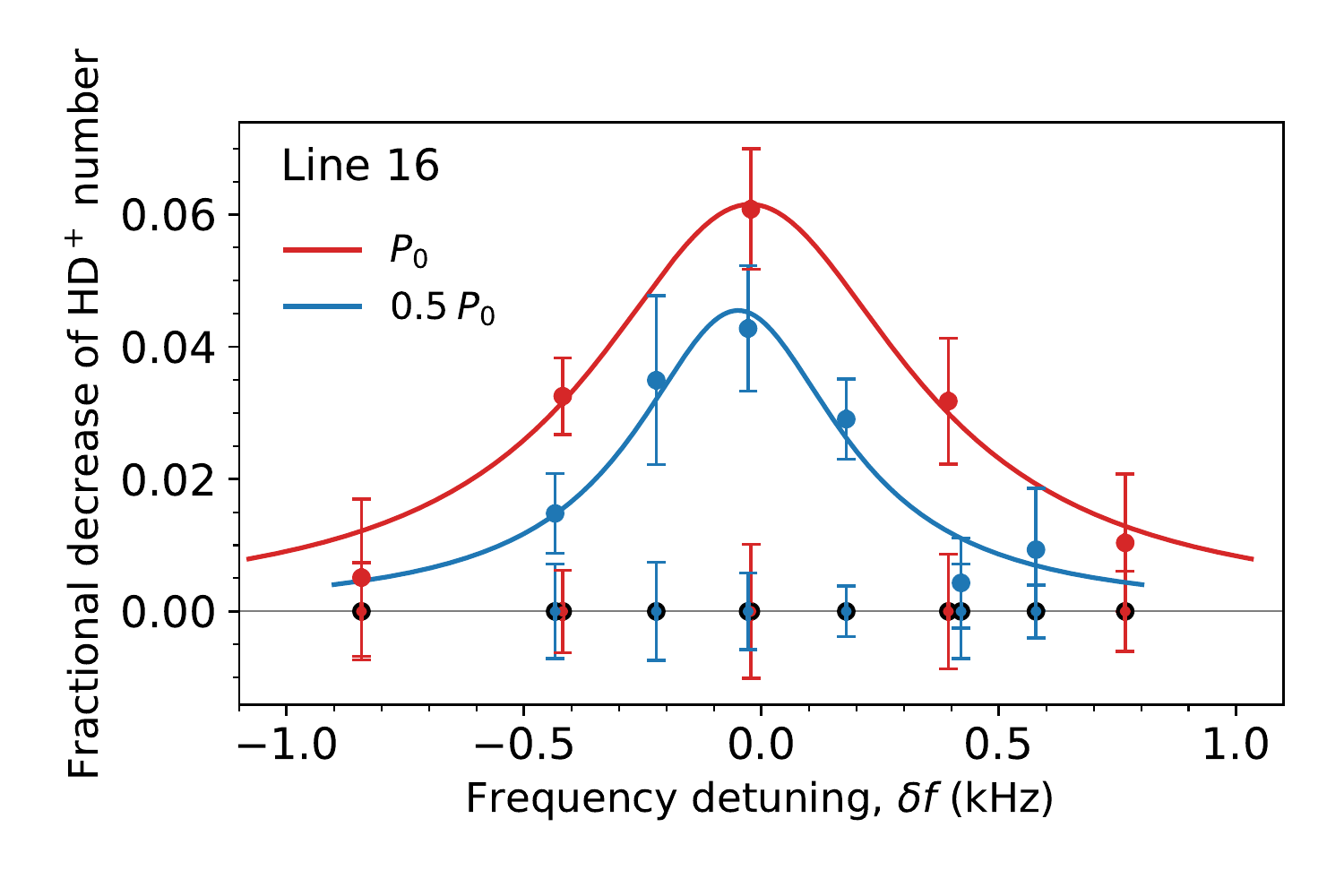}}
\par\end{center}

\noindent Supplementary Figure 2. \textbf{Power broadening of one Zeeman component of the vibrational
transition.} The $m_{F}=0\rightarrow m_{F}'=0$ component of line
16 was interrogated at two power settings of the $5.1\,\mu{\rm m}$
spectroscopy laser. Here, $B\simeq0.6\,{\rm G}$. The zero of the
frequency detuning is arbitrary. $P_{0}$ is the nominal power used
in the measurements shown in the main text. Each error bar represents the standard deviation of the mean.

\clearpage

\begin{center}
\textbf{\includegraphics[width=\columnwidth]{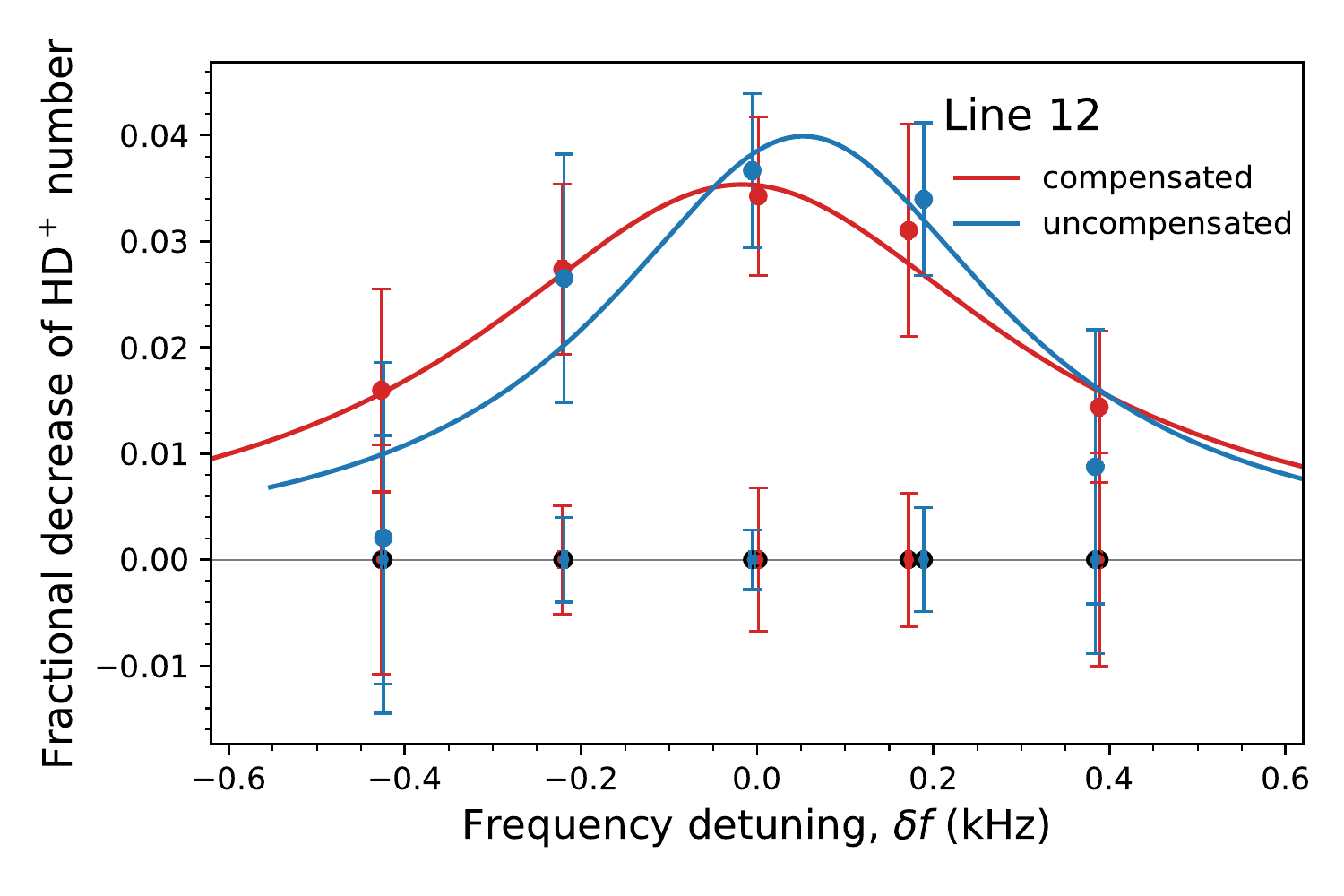}}
\par\end{center}

\noindent Supplementary Figure 3. \textbf{The effect of a displacement of the beryllium ion cluster
from the trap axis.} One Zeeman component of the vibrational transition
is measured with the beryllium ion cluster aligned with the trap axis (red) and with the cluster significantly shifted radially (blue). The $m_{F}=0\rightarrow m_{F}'=0$
component of line 12 is shown. The zero of the frequency detuning
is arbitrary. Each error bar represents the standard deviation of the mean.

\end{NoHyper}